\documentclass[12pt]{article}
\usepackage{epsfig}
\begin{document}
\title{\bf New bound on penguin pollution in $B^0(t) \to \pi^+ \pi^-$ }
\author{Anjan K. Giri$^1$ and Rukmani Mohanta$^2$\\
\\
$^1$ {\it Physics Department, Panjab University,}\\
{Chandigarh-160014, India}\\
$^2$ {\it School of Physics, University of Hyderabad,}\\
{Hyderabad-500 046, India} } \maketitle
\begin{abstract}
In the presence of penguin contributions, the indirect CP
asymmetry in $B^0(t) \to \pi^+ \pi^-$ measures $\sin 2\alpha_{eff}$,
where $\alpha_{eff}$ deviates from the true value of the CKM phase
$\alpha$ by an amount $\theta$ i.e., $\alpha_{eff}=\alpha  + \theta $.
Using the measured value of direct CP asymmetry in $B^0 (t) \to
\pi^+ \pi^- $, we derive  new  bound on $|2 \theta|$.
\end{abstract}

The study of CP violation mechanism is one of the main goals of
the ongoing and future B factory experiments \cite{ref1}. In the
standard model (SM) CP violation is induced by the nonzero phase
appearing in the CKM mixing matrix and is often characterized by
the so called unitarity triangle \cite{ref1a}. Detection of CP
violation and the accurate determination of the unitarity triangle
are the major goals of experimental B Physics. Decisive
information about the origin of CP violation in the flavor sector
can be obtained  if the three angles $\alpha$, $\beta $ and
$\gamma$ of the unitarity triangle can be independently measured.
The sum of these three angles must be equal to $180^\circ $, if
the CKM phenomena of SM is the model for CP violation. The usual
way to measure the three angles of the unitarity triangle 
is generally done by considering the time dependent rate
asymmetries of  B meson decays \cite{ref1b}.

The angle $\beta $ is the simplest one among these three
angles, which can be determined from the time dependent rate asymmetry
of the gold plated mode $B_d^0 \to J/\psi K_S$, without
any uncertainty. In fact the
value of $\sin(2 \beta)$ has recently been reported
by the Belle \cite{ref2} and BaBar \cite{ref3}
collaborations as

\begin{eqnarray}
\sin ( 2\beta) &=& 0.99 \pm 0.14~ (stat.) \pm 0.06 ~(syst.)~~~~~~
({\rm Belle})\nonumber\\
 \sin (2\beta) &=& 0.75 \pm 0.09~ (stat.)
\pm 0.04~ (syst.)~~~~~~ ({\rm BaBar})
\end{eqnarray}
with an average
\begin{equation}
\sin (2 \beta) = 0.87 \pm 0.08
\end{equation}

The most difficult among these angles is the angle $\gamma $. But
there are several methods exist  for its  determination
\cite{ref4,ref5,ref5a}. Recently, it has also been pointed out
that the decay modes  $B_c \to D_s D^0$,
 $B_s \to D^0 \phi$  and $ \Lambda_b \to \Lambda D^0$ \cite{ref6}
can be used to cleanly determine the value
of $\gamma $.

For the measurement of $\alpha $, the principal decay mode
considered is $ B^0(t) \to \pi^+ \pi^-$. However, due to the
penguin pollution, a clean determination of the angle $\alpha$ is
not possible considering this decay mode \cite{ref7}. The problem
of penguin pollution can be eliminated with the help of isospin
analysis \cite{ref8}. In order to perform the isospin analysis, all
modes of $B \to \pi \pi$ have to be measured. However, it is
difficult to measure the branching ratio of $B^0(\bar B^0) \to
\pi^0 \pi^0$ process, which has background problem as well as a
tiny branching ratio of ${\cal O}(10^{-7})$ \cite{ref8a}.
Therefore, in practice it is hard to perform the isospin analysis.

To be more precise, in the presence of penguin contributions, the
CP asymmetry in $B^0(t) \to \pi^+ \pi^-$ does not measure $\sin
(2\alpha)$ but $\sin (2 \alpha+2\theta )$, where $2 \theta $
parameterizes the effect of penguin contributions. There are
several methods exist in the literature to constrain the penguin
pollution parameter $\theta $. Grossman and Quinn \cite{ref9} have
shown that an upper bound on the error on $\sin(2 \alpha)$ due to
penguin pollution can be obtained using only the measured rate
$Br(B^\pm \to \pi^\pm \pi^0)$ and an upper bound on the combined
rate $Br(B^0 \to \pi^0 \pi^0)+Br(\bar B^0 \to \pi^0 \pi^0)$.
Later, Charles \cite{ref10} improved the bound to some extent. In
addition the paper \cite{ref10} presents several bounds based on
the flavor SU(3) symmetry, together with dynamical assumptions
about the size of certain OZI-suppressed penguin contributions.
The recent one is being proposed  by Gronau et al \cite{ref11}.
Although the bound is an improvement over the last two bounds but
it depends on the measurement on $Br(B^0 \to \pi^0 \pi^0) $, which
posses a serious challenge for the experimentalists. Therefore,
the bound given by Gronau et al may not solve the purpose
immediately. This in turn necessitates further scrutiny. Since
with the accumulation of more and more data samples day by day the
Belle and BaBar will be in a position to analyze $B \to \pi \pi$
mode more precisely in the near future. This raises a question
whether there exists any other way to constrain the penguin
parameter $(\theta)$ exploring the experimental data, as of today.

Our effort in this paper will be an attempt in this direction to
obtain some bound on penguin induced shift $|2\theta|$. In
contrast to the earlier bounds, we show that the measurement of
$B^0(t) \to \pi^+ \pi^-$ observables can be used to put a lower
limit on $\cos 2 \theta $.

We now present a new bound in terms of the direct CP asymmetry
$a_{dir}$ of $B^0(t) \to \pi^+ \pi^-$, which has already
been measured at KEK \cite{ref12} and SLAC {\cite{ref13}.
So it may be worthwhile  to put some
constraint on the penguin pollution in $B \to \pi \pi$ in terms
of $a_{dir}$, which may help us for
the extraction of $\alpha $.

We first consider the decay mode $B^0 \to \pi^+ \pi^-$.
As it is done usually, defining
$T$ as the color favored tree amplitude,
and $P$ as the penguin amplitude, the transition amplitude for
$B^0 \to \pi^+ \pi^-$ and  $\bar {B^0} \to \pi^+ \pi^-$ modes are
given as

\begin{eqnarray}
A^{+-} &\equiv & A( B^0 \to \pi^+ \pi^-)= Te^{+ i \gamma} +
Pe^{-i \beta}=\nonumber\\
&=&e^{+i \gamma }\left [ T - P e^{+i \alpha } \right ]
\end{eqnarray}
\begin{eqnarray}
\bar A^{+-} &\equiv & A( {\bar B^0} \to \pi^+ \pi^-)= Te^{- i \gamma} +
Pe^{+i \beta}\nonumber\\
&=&e^{-i \gamma }\left [ T - P e^{-i \alpha } \right ]
\end{eqnarray}
where we have used the Wolfenstein parameterization of CKM matrix
and substituted $\alpha= \pi -(\beta +\gamma)$.

The time dependent rate asymmetry in $B^0(t) \to \pi^+ \pi^-$
is given as
\begin{eqnarray}
a_{CP}(t)&=& \frac{\Gamma(B^0(t) \to \pi^+ \pi^-) -\Gamma(\bar
B^0(t) \to \pi^+ \pi^-)} {\Gamma(B^0(t) \to \pi^+ \pi^-)
+\Gamma(\bar B^0(t) \to \pi^+ \pi^-)}
\nonumber\\
&=&a_{dir} \cos \Delta m t +a_{mix-ind} \sin \Delta m t
\end{eqnarray}
where
\begin{equation}
a_{dir}\equiv \frac{1-|\lambda _{\pi^+ \pi^-}|^2} {1+|\lambda
_{\pi^+ \pi^-}|^2}\;,~~~~~~~~~~~~ a_{mix-ind}\equiv \frac{-2 {\rm
Im}(\lambda _{\pi^+ \pi^-})} {1+|\lambda _{\pi^+
\pi^-}|^2}\;,~~~~~~~~~~~~
\end{equation}
with
\begin{equation}
\lambda _{\pi^+ \pi^-}\equiv e^{-2i \beta}~\frac{A( \bar{B^0} \to
\pi^+ \pi^-)}{A( B^0 \to \pi^+ \pi^-)}
\end{equation}

We now define the average branching ratio ${\cal B}_{\pi^+ \pi^-}$
for the decay mode $B^0 \to \pi^+ \pi^-$  and $\bar B^0 \to \pi^+ \pi^-$
 as
\begin{eqnarray}
{\cal B}_{\pi^+ \pi^-}&=& \frac{1}{2}\left [
Br(B^0 \to \pi^+ \pi^-)+Br( \bar B^0 \to \pi^+ \pi^-) \right ]
\nonumber\\
&\equiv & \frac{1}{2}\left [ |A^{+-}|^2 +|\bar A^{+-}|^2 \right ]
\end{eqnarray}
where we express the amplitudes squared in units of two body
branching ratios. Substituting the value of $ \lambda_{\pi^+
\pi^-}$ from Eq. (7), we obtain the expression for direct CP asymmetry
$(a_{dir})$ as
\begin{equation}
a_{dir} = \frac{|A^{+-}|^2 -|\bar A^{+-}|^2}
{|A^{+-}|^2 +|\bar A^{+-}|^2}
\end{equation}

Recently Belle Collaboration \cite{ref12} reported their first measurement
of the CP violating parameters in $B^0 \to \pi^+ \pi^-$ decay
\begin{eqnarray}
&&a_{mix-ind} = -1.21_{-0.27}^{+0.38}~(stat.)_{-0.13}^{+0.16}~(syst.)\nonumber\\
&&a_{dir} = 0.94_{-0.31}^{+0.25} ~(stat.)\pm 0.09 ~(syst.)\;.
\end{eqnarray}
This is in comparison to the previous BaBar result \cite{ref13}
\begin{eqnarray}
&& a_{mix-ind} =- 0.01 \pm 0.37~ (stat.)\pm 0.07~ (syst.) \nonumber\\
&& a_{dir} = 0.02\pm 0.29~ (stat.)\pm 0.07~ (syst.)
\end{eqnarray}

Taking into account both  the Belle and BaBar measurements, the
average of $a_{mix-ind}$ and $a_{dir}$ is given as
\begin{equation}
a_{mix-ind} = -0.64 \pm 0.26\;,~~~~~~~~~~~~ a_{dir} = 0.49 \pm
0.21
\end{equation}
One can easily see from Eqs. (6) and (7) that,
if we neglect the penguin contribution than
we have $a_{dir}=0$ and $a_{mix-ind}=\sin 2 \alpha$. That means we
can measure $\sin 2 \alpha $ directly from $B^0(t) \to \pi^+ \pi^-$
decay. However, due to the presence of penguin contributions
the extracted value of $\alpha_{eff} $ from $B^0 (t) \to \pi^+ \pi^-$
deviates from the true $\alpha $ value. We define $\alpha_{eff}$
as
\begin{equation}
2 \alpha_{eff} ={\rm Arg}\left [e^{-2 i \beta} \bar A^{+-} {A^{+-}}^*
\right ]
\end{equation}

Thus it can be seen that
the observables in $B^0(t) \to \pi^+ \pi^-$ are the average
branching ratio and the CP asymmetries :
\begin{equation}
{\cal B}_{\pi^+ \pi^-},~a_{dir} ~~{\rm and}~~a_{mix-ind} \equiv
\sin (2 \alpha_{eff})
\end{equation}
As we have already noted that the vanishing of penguin amplitude
$P$ implies $2 \alpha_{eff} =2 \alpha$. Thus the magnitude of
penguin amplitude can be expressed as some function of $2 \theta $
(say), where $2 \theta = (2\alpha_{eff}-2 \alpha)$.

From Eqs. (3) and (4) we can write
\begin{equation}
(2i \sin \alpha ) P = e^{-i \gamma} A^{+-} - e^{+i \gamma}
\bar A^{+-}
\end{equation}
Substituting the values of ${\cal B}$ and $a_{dir}$ we can
express the squared magnitude of penguin contribution as
\begin{equation}
|P|^2 =\frac{{\cal B}_{\pi^+ \pi^-}}{1-\cos 2 \alpha}
\left [1- \sqrt{1-a_{dir}^2}\cos 2 \theta \right ]
\end{equation}

Now we consider the three $B \to \pi \pi $ decay amplitudes as

\begin{eqnarray}
A^{+-} &\equiv & A(B^0 \to \pi^+ \pi^-) = Te^{i \gamma}+Pe^{-i
\beta} =\left (T-P e^{i \alpha} \right ) e^{i \gamma}
\nonumber\\
A^{00} &\equiv &  A(B^0 \to \pi^0 \pi^0) = \frac{1}{\sqrt 2} \left
( Ce^{i \gamma}-Pe^{-i \beta}\right ) =\frac{1}{\sqrt 2} \left
(C+P e^{i \alpha} \right )e^{i \gamma}
\nonumber\\
A^{+0} &\equiv &  A(B^+ \to \pi^+ \pi^0) =
\frac{1}{\sqrt 2} \left ( C +T \right )e^{i \gamma}
\end{eqnarray}
where the CP conserving  complex amplitudes $T$, $C$ and $P$ denote the
`tree', `color suppressed' and `penguin' contributions. Here
we have neglected the small electroweak penguin contributions
in the decay mode $B^+ \to \pi^+ \pi^0$. These amplitudes obey
the isospin triangle relation
\begin{equation}
\frac{1}{\sqrt 2} A^{+-} +A^{00}=A^{+0}
\end{equation}
The corresponding $\bar A $ amplitudes can be obtained from the
$A$ amplitudes by simply changing the signs of weak phases.

Now we define the new amplitudes as
\begin{equation}
B^{ij}=e^{2i \gamma} \bar A^{ij}\;.
\end{equation}
It is obvious that $|B^{ij}|=|\bar A^{ij}|$. Thus we can
explicitly write the $B$ amplitudes as
\begin{eqnarray}
B^{+-} & = &\left (T-Pe^{-i \alpha} \right )e^{i \gamma}
\nonumber\\
B^{00} & = &
\frac{1}{\sqrt 2} \left ( C+Pe^{-i \alpha}\right ) e^{i \gamma}
\nonumber\\
B^{-0} & = &
\frac{1}{\sqrt 2} \left ( C +T \right )e^{i \gamma}
\end{eqnarray}
Similar to (18) the $B$ amplitudes also obey
the isospin triangle relation
\begin{equation}
\frac{1}{\sqrt 2} B^{+-} +B^{00}=B^{-0}
\end{equation}

It should be noted that $A^{+0}=B^{-0}$, so that
$A$ and $B$ isospin triangles have a common base. Secondly,
in the absence of penguin contributions $B^{+-}=A^{+-}$.
Thus the relative phase $2 \theta $  between these two amplitudes
is due to penguin pollution. These two isospin triangles are
depicted in Figure-1. One can easily note that the distance
between the points $E$ and $H$ (i.e. the difference between
$A^{+-}/\sqrt 2$ and $B^{+-}/\sqrt 2$ ) is
\begin{equation}
|EH|={\sqrt 2}|P| \sin \alpha
\end{equation}
We now consider the triangle EFH with the interior angles as
$\angle EFH = 2 \theta $ and $\angle EHF =\theta_1$.
Using the sine theorem we can write
\begin{equation}
\sin 2 \theta =\sin\theta_1 ~\frac{{\sqrt 2}
|P| \sin \alpha}{\frac{1}{\sqrt 2}
~|A^{+-}|}
\end{equation}
We impose now the boundary condition on the angle $\theta_1$. Since
the triangle is closed, $\theta_1$ must lie in the range
$0<\theta_1< 180^\circ $, which implies that
$0< \sin \theta_1 \leq 1$, the maximum value is being for
$\theta_1=90^\circ $. Thus we can write

\begin{equation}
\sin 2 \theta \leq  \frac{2|P| \sin \alpha}{
|A^{+-}|}
\end{equation}
Squaring  both sides of the above inequality and substituting the
value of $4|P|^2 \sin^2 \alpha $ from Eq. (16) and
$|A^{+-}|^2={\cal B}_{\pi^+ \pi^-}(1+a_{dir})$, one can obtain
from (24)
\begin{equation}
\sin^2 2 \theta \leq \frac{ 2 \left [1- \sqrt{1-a_{dir}^2} \cos 2 \theta
\right ]}{1+a_{dir}}
\end{equation}
After simplification the above inequality can be given as

\begin{equation}
\left |\sqrt{1+a_{dir}} \cos 2 \theta - \sqrt{1-
a_{dir}} \right | \geq 0
\end{equation}
The inequality in (26) does not provide any information regarding
the relative signs between the two terms. That is we can not draw
any conclusion from the above equation whether the first term is
greater than or less than the second term and hence no bound on
$\cos2 \theta $ can be obtained. Therefore, to find out the bound
on $\cos 2 \theta $ we have to consider other informations as
well. It should be noted that the equality sign holds in Eq. (25)
only for $\theta_1=90^\circ $. This allows us to write the
inequality (25) as

\begin{equation}
\frac{ 2 }{1+a_{dir}} \left [1- \sqrt{1-a_{dir}^2} \cos 2 \theta
\right] < 1 \;,
\end{equation}
which gives us

\begin{equation}
\cos 2 \theta > \frac{1}{2} \sqrt{\frac{1-a_{dir}}{1+ a_{dir}}}\;.
\end{equation}

Now let us assume that $2\theta  < 90^\circ $, then considering
the right angled triangle EIF with $\angle EIF = 90^\circ $, we
can write
\begin{equation}
\cos 2 \theta = \frac{|FI|}{|EF|} \leq \frac{|FH|}{|EF|}
\end{equation}
Again the equality sign is for $\theta_1=90^\circ $. Thus we obtain

\begin{equation}
\cos 2 \theta \leq \frac{|B^{+-}|}{|A^{+-}|} = \frac{|B^{+-}|
|A^{+-}|}{|A^{+-}|^2}
\end{equation}
It should be noted that $|B^{+-}|=|\bar A^{+-}|$. Now substituting
$|\bar A^{+-} A^{+-}| ={\cal B}_{\pi^+ \pi^-}\sqrt{1-a_{dir}^2}$
and $|A^{+-}|^2={\cal B}_{\pi^+ \pi^-}(1+a_{dir})$ we obtain the
bound on $\cos 2 \theta $ as

\begin{equation}
\cos 2 \theta \leq\sqrt{\frac{1-a_{dir}}{1+a_{dir}}}
\end{equation}

The above inequality is valid only when $2\theta < 90^\circ$.
Combining (28) and (31)  the bound on $\cos 2 \theta $ is given
(subject to above restriction) as

\begin{equation}
\frac{1}{2}\sqrt{\frac{1-a_{dir}}{1+a_{dir}}} < \cos 2 \theta
\leq\sqrt{\frac{1-a_{dir}}{1+a_{dir}}}
\end{equation}

This is the new  bound on $\cos 2 \theta$.
Here the bound on $\cos 2 \theta $ is given in terms
of the measurable quantity $a_{dir}$ only.
It should be noted here that this bound has been derived
assuming that the isospin triangles are closed and they have a
common base. Thus, to the extent that isospin is violated,
whether by electroweak penguin contribution or by $\pi^0-\eta$,
$\eta^\prime $ mixing \cite{ref14}, the bound will be correspondingly
weakened.

Substituting the current average value of $a_{dir}$ from Eq. (12)
the bound on $\cos 2 \theta $ is given as
\begin{equation}
0.2925 < \cos 2 \theta \leq 0.585.
\end{equation}

To summarize, in this paper we have derived a new upper bound on
the penguin induced shift $|2 \theta |$ in $B^0 \to \pi^+ \pi^-$
decay. In contrast to the earlier bounds, we have shown here that
the measurement of direct CP asymmetry in $B^0(t) \to \pi^+ \pi^-
$ can be used to place some limit on the penguin pollution
parameter.

We would like to thank Professor David London for useful comments.
One of us (AKG) would like to thank Council of Scientific and Industrial
Research, Government of India, for financial support.

\begin{figure}
\centerline{\epsfysize=10in\epsffile{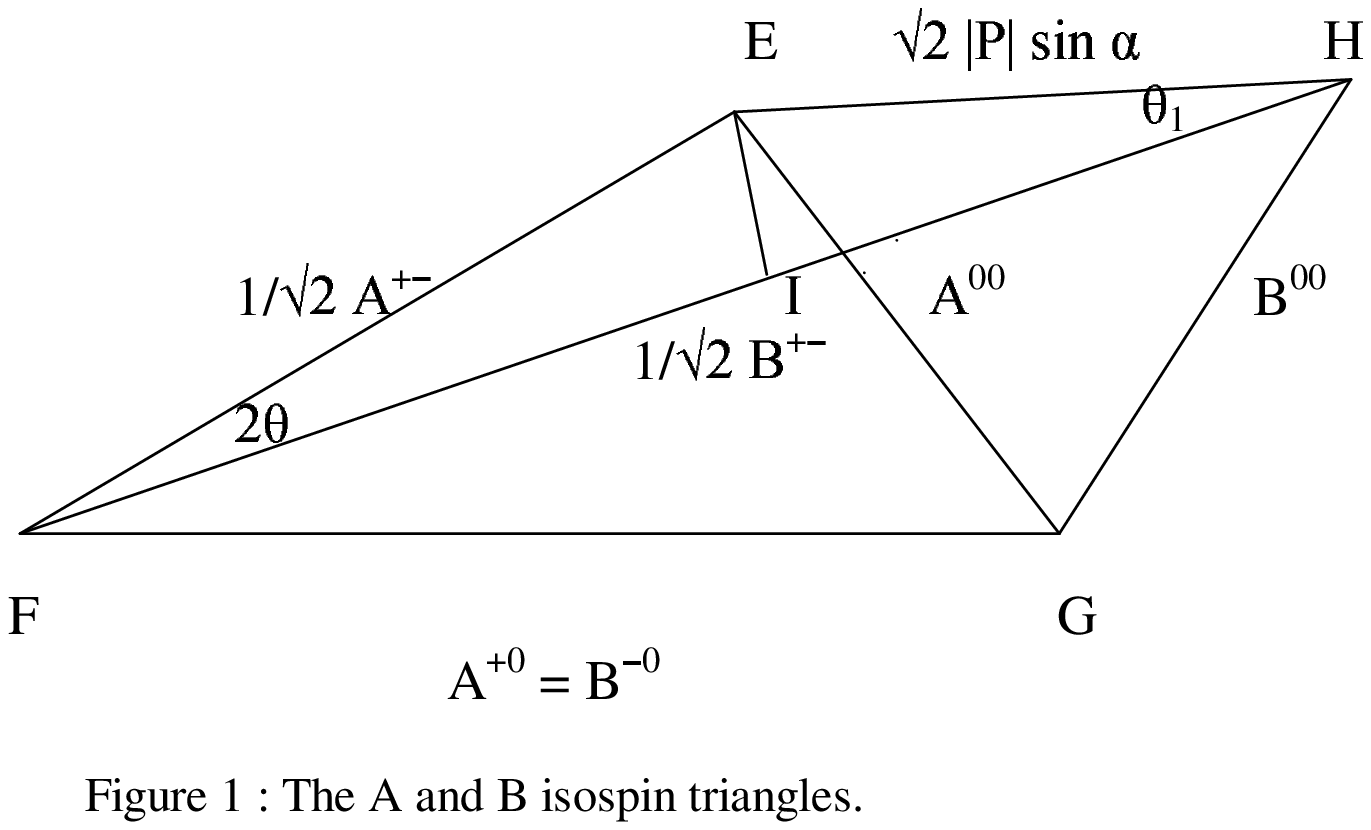}}
\label{}
\end{figure}


\begin{thebibliography}{99}

\bibitem{ref1} For a review, see {\it B Decays}, edited by S. Stone,
(World Scientific, Singapore, 1994).
\bibitem{ref1a} L. L. Chau and W. -Y. Keung, Phys. Rev. Lett.
{\bf 53}, 1802 (1984); C. Jarlskog and R. Stora, Phys. Lett. {\bf
B 208}, 268 (1988).
\bibitem{ref1b} I. I. Bigi and A. I. Sanda, Nucl. Phys. {\bf B 193},
85 (1981).
\bibitem{ref2} K. Abe et al. Belle Collaboration, hep-ex/0202027.
\bibitem{ref3} B. Aubert et al. BaBar Collaboration, hep-ex/0203007.
\bibitem{ref4} M. Gronau and D. London, Phys. Lett. {\bf B  253},
483 (1991); M. Gronau and D. Wyler, Phys. Lett. {\bf B 265},
172 (1991).
\bibitem{ref5} R. Aleksan, I. Dunietz and B. Kayser, Z. Phys. {\bf C
54}, 653 (1992).
\bibitem{ref5a} M. Gronau, J. L. Rosner and D. London, Phys. Rev.
Lett. {\bf 73}, 21 (1994); R. Fleischer, Phys. Lett. {\bf B 365},
399 (1996); M. Neubert and J. L. Rosner, Phys. Rev. Lett. {\bf 81},
5076 (1998).
\bibitem{ref6} A. K. Giri, R. Mohanta and M. P. Khanna, Phys. Rev.
{\bf D 65}, 034016 (2002); Phys. Rev.
{\bf D 65}, 056015 (2002); Phys. Rev.
{\bf D 65}, 073029 (2002).
\bibitem{ref7} D. London, R. Peccei, Phys. Lett. {\bf B 223}, 257
(1989); M. Gronau, Phys. Rev. Lett {\bf 63}, 1451 (1989); Phys. Lett.
{\bf B 300}, 163 (1993); B. Grinstein Phys. Lett. {\bf B 229}, 280 (1989).

\bibitem{ref8} M. Gronau and D. London, Phys. Rev. Lett. {\bf 65},
3381 (1990).
\bibitem{ref8a} C. D. L\"u, K. Ukai and M.-Z. Yang,
Phys. Rev. {\bf D 63}, 074009 (2001).
\bibitem{ref9} Y. Grossman and H. R. Quinn, Phys. Rev. {\bf D 58},
017504 (1998).
\bibitem{ref10} J. Charles, Phys. Rev. {\bf D 59}, 054007 (1999).
\bibitem{ref11} M. Gronau, D. London, N. Sinha and R. Sinha,
Phys. Lett. {\bf B 514}, 315 (2001).
\bibitem{ref12} K. Abe et al. Belle Collaboration, hep-ex/0204002.
\bibitem{ref13} B. Aubert et al. BaBar Collaboration,
hep-ex/0205082.
\bibitem{ref14} S. Gardner, Phys. Rev. {\bf D 59}, 077502 (1999).

\end{thebibliography}
\end{document}